# Weakening Assumptions for Deterministic Subexponential Time Non-Singular Matrix Completion

Maurice Jansen[*]

Institute for Theoretical Computer Science
Tsinghua University, Beijing, China.
mjjansen@tsinghua.edu.cn.

**Abstract**

Kabanets and Impagliazzo [KI04] show how to decide the circuit polynomial identity testing problem (CPIT) in deterministic subexponential time, assuming hardness of some explicit multilinear polynomial family $\{f_m\}_{m \geq 1}$ for arithmetic circuits.

In this paper, a special case of CPIT is considered, namely non-singular matrix completion (NSMC) under a low-individual-degree promise. For this subclass of problems it is shown how to obtain the same deterministic time bound, using a weaker assumption in terms of the *determinantal complexity* $\mathrm{dc}(f_m)$ of $f_m$.

Hardness-randomness tradeoffs will also be shown in the converse direction, in an effort to make progress on Valiant's VP versus VNP problem. To separate VP and VNP, it is known to be sufficient to prove that the determinantal complexity of the $m \times m$ permanent is $m^{\omega(\log m)}$. In this paper it is shown, for an appropriate notion of explicitness, that the existence of an explicit multilinear polynomial family $\{f_m\}_{m \geq 1}$ with $\mathrm{dc}(f_m) = m^{\omega(\log m)}$ is equivalent to the existence of an efficiently computable *generator* $\{G_n\}_{n \geq 1}$ *for* multilinear NSMC with seed length $O(n^{1/\sqrt{\log n}})$. The latter is a combinatorial object that provides an efficient deterministic black-box algorithm for NSMC. "Multilinear NSMC" indicates that $G_n$ only has to work for matrices $M(x)$ of $poly(n)$ size in $n$ variables, for which $\det(M(x))$ is a multilinear polynomial.

## 1 Introduction

Let $\mathbb{F}$ be a field of characteristic zero, let $\mathbb{Q} \subseteq \mathbb{F}$ denote the field of rational numbers, and let $X_n = \{x_1, x_2, \ldots, x_n\}$ be a set of variables. $\mathcal{A}_{\mathbb{F}}(X_n)$ denotes the set of affine linear forms over $X_n$ and $\mathbb{F}$. In this paper we study a special case of circuit polynomial identity testing, namely the non-singular matrix completion problem over $\mathbb{F}$. Matrix completion is an important problem, both in theory and in practice. The history of the problem dates back to work by Lovász [Lov79] and Edmonds [Edm67].

As was done in [DSY08] for CPIT, we study non-singular matrix completion under a promise restriction on individual degrees:

[*]This work was supported in part by the National Natural Science Foundation of China Grant 60553001, and the National Basic Research Program of China Grant 2007CB807900,2007CB807901.

**Problem** $\text{NSMC}_r^k(\mathbb{F})$ : $k \times k$ Non-Singular Matrix Completion over $\mathbb{F}$ with individual degrees at most $r$.

- Input: A $k \times k$ matrix $M(x)$ with entries in $\mathcal{A}_{\mathbb{F}}(X_n)$.

- Promise : Individual degrees of the polynomial $\det(M)$ are bounded by r.

- Question: Does there exist $a \in \mathbb{F}^n$ such that $\det M(a) \neq 0$ ?

Over a field of characteristic zero, the problem is equivalent to asking whether $\det M(x) \not\equiv 0$. Since $\det_n$ has $O(n^6)$ size skew circuits [MV97], and is universal for skew circuits (Implicit in [Val79], see Proposition 3.1), $\text{NSMC}_{r(n)}^{poly(n)}(\mathbb{F})$ is equivalent to identity testing $poly(n)$ size skew circuits over $\mathbb{F}$, under the *semantic* promise that the circuit outputs a polynomial with individual degrees bounded by $r(n)$. Over $\mathbb{Q}$, for any $r(n)$, the latter can be verified with a coRP-algorithm, using the Schwartz-Zippel Lemma [Sch80, Zip79]. Moreover, Lovász showed that a random assignment for $x$ maximizes the rank of $M(x)$ with high probability [Lov79].

Whether there exists an efficient deterministic algorithm for matrix completion is a major open problem. Currently, such an algorithm exists only for special instances. For example, Ivanyos, Karpinkski and Saxena give a polynomial time deterministic algorithm for finding a maximum rank completion, provided $M(x)$ is of the form $M_0 + x_1 M_1 + x_2 M_2 + \ldots + x_n M_n$, where $M_1, M_2, \ldots, M_n$ are rank one matrices [GI09].

Kabanets and Impagliazzo provide algebraic hardness-randomness tradeoffs for CPIT [KI04]. They show that the existence of an explicit polynomial with super-polynomial arithmetic circuit size, implies CPIT, and hence NSMC, can be decided deterministically in time $2^{n^\epsilon}$, for any $\epsilon > 0$, provided $n$ is large enough. In order to make progress towards unconditionally proven deterministic subexponential time algorithms for NSMC, it is important to consider whether the same bound can be obtained for NSMC under any weaker assumptions.

In this paper we will only assume hardness of an explicit polynomial for skew circuits, or equivalently, we make hardness assumptions in terms of *determinantal complexity* [MR04]. In other words, we aim for specialized algebraic hardness-randomness tradeoffs for the skew circuit model. For this, we will use the hardness-randomness tradeoffs for constant-depth arithmetic circuits due to Dvir, Shpilka and Yehudayoff [DSY08] as a starting point.

Another motivation is the VP versus VNP question, or the permanent versus determinant problem [MR04]. The latter problem asks of us to prove lower bounds for the determinantal complexity of an explicit[1] polynomial. We firmly establish the role of NSMC in the quest for such lower bounds, firstly, by the characterization mentioned in the abstract. Secondly, it is shown that the existence of an explicit multilinear polynomial family $\{f_m\}_{m \geq 1}$ with $\text{dc}(f_m) = m^{\omega(1)}$ is equivalent to the existence of an efficiently computable multilinear generator $\{G_n\}_{n \geq 1}$ for $\text{NSMC}_1^{poly(n)}$ with seed length $\lceil n^\epsilon \rceil$, for some $0 < \epsilon < 1$.

## 2 Results

We require some formal definitions to properly state the results.

---

[1] Necessarily in the sense of Definition 2.2. A sufficient condition would require an even more stringent notion.

**Definition 2.1 ([MR04])** *The* determinantal complexity $dc(f)$ *of a polynomial* $f \in \mathbb{F}[X_n]$ *is defined to be the minimum size of a matrix $M$ with entries in $\mathcal{A}_\mathbb{F}(X_n)$ such that* $\det M = f$.

We use standard definitions of arithmetic circuits with binary addition and multiplication operations (See [BCS97]). Arithmetic circuit complexity of $f$ is denoted by $L(f)$. A *skew* circuit satisfies that for every multiplication gate one of its inputs is a variable or a constant. $L_{skew}(f)$ denotes skew circuit size of $f$. The following is our notion of explicitness of a multilinear polynomial:

**Definition 2.2** *Let $\{f_m\}_{m \geq 1}$ be a family of multilinear polynomials with $f_m \in \mathbb{Z}[x_1, x_2, \ldots, x_m]$. We say this family is* explicit *provided there exists a deterministic Turing machine running in time $2^{O(m)}$, that on input $e \in \{0,1\}^m$, outputs the binary representation of the coefficient of the monomial $x_1^{e_1} x_2^{e_2} \ldots x_m^{e_m}$ of $f_m$.*

**Hardness Hypothesis 1 (HH1)** *There exists an explicit family of multilinear polynomials $\{f_m\}_{m \geq 1}$, such that $L(f_m) = m^{\omega(1)}$.*

**Hardness Hypothesis 2 (HH2)** *There exists an explicit family of multilinear polynomials $\{f_m\}_{m \geq 1}$, such that $dc(f_m) = m^{\omega(1)}$.*

If in the above we replace $m^{\omega(1)}$ by $m^{\omega(\log m)}$, we refer to this as Strengthened HH1 and Strengthened HH2.

**Proposition 2.3** *HH2 is equivalent to the statement that there exists an explicit family of multilinear polynomials $\{f_m\}_{m \geq 1}$, such that $L_{skew}(f_m) = m^{\omega(1)}$. A similar statement holds for Strengthened HH2, but with $m^{\omega(1)}$ replaced by $m^{\omega(\log m)}$.*

**Proof.** In one direction this follows from the fact that the $n \times n$ determinant has skew circuits of size $O(n^6)$ [MV97]. For the converse, apply the fact that if $f_m$ can be computed by a skew circuit of size $s$, then $dc(f_m) = O(s)$ (Implicit in [Val79], see Proposition 3.1). $\square$

**Proposition 2.4** *Strengthened HH1 $\Rightarrow$ Strengthened HH2 $\Rightarrow$ HH1 $\Rightarrow$ HH2.*

**Proof.** The first and the last implication follow from Proposition 2.3. To show that Strengthened HH2 $\Rightarrow$ HH1, suppose we have an explicit multilinear $p$-family $\{f_m\}_{m \geq 1}$, such that $dc(f_m) = m^{\omega(\log m)}$. This implies $dc(f_m) = m^{\omega(\log m)}$, even when restricting to $m \in \mathcal{M}$, for any infinite set $\mathcal{M}$. If $L(f_m) \notin m^{\omega(1)}$, then there exists constant $c > 0$ and infinite set $\mathcal{M}'$, such that $L(f_m) \leq m^c$, for all $m \in \mathcal{M}'$. Using the construction of [VSBR83], we obtain formulas for $f_m$ of size $2^{O(\log L(f_m) \log m)} = m^{O(\log m)}$, for $m \in \mathcal{M}'$. Hence by [Val79], $dc(f_m) = m^{O(\log m)}$, for $m \in \mathcal{M}'$. This is a contradiction. $\square$

Our algorithms will be of the black-box kind. This is formalized as follows:

**Definition 2.5** *For a function $\ell : \mathbb{N} \to \mathbb{N}$, a multilinear $(\ell(n), n)$-generator for $\mathrm{NSMC}_r^k(\mathbb{F})$ is given by a multilinear polynomial mapping $G_n : \mathbb{F}^{\ell(n)} \to \mathbb{F}^n$. We say $G_n$ provides a test for $\mathrm{NSMC}_r^k(\mathbb{F})$, if for any instance $M(x)$ of $\mathrm{NSMC}_r^k(\mathbb{F})$, it holds that*

$$(\exists a \in \mathbb{F}^n), \det M(a) \neq 0 \text{ iff } (\exists b \in \mathbb{F}^{\ell(n)}), \det M(G_n(b)) \neq 0.$$

Families $\{G_n\}_{n \geq 1}$ of generators are also simply called "generator". For a generator $\{G_n\}_{n \geq 1}$ with coefficients in $\mathbb{Z}$, we say it is *efficiently computable*, if there exists a deterministic Turing machine $M$ that runs in time $2^{O(\ell(n))}$, so that on input $(i, n, e)$, where $i$ and $n$ are given in binary and $e \in \{0, 1\}^{\ell(n)}$, $M$ computes the binary representation of the coefficient of the monomial $x_1^{e_1} x_2^{e_2} \ldots x_{\ell(n)}^{e_{\ell(n)}}$ of $(G_n)_i$.

We are now ready to state the results.

**Theorem 2.6** *If HH2 holds over $\mathbb{F}$, then for any $0 < \epsilon < 1$, there exists an efficiently computable multilinear $(\lceil n^\epsilon \rceil, n)$-generator $\{G_n\}_{n \geq 1}$, such that for any $k(n) \in n^{O(1)}$ and $r(n) \in O(1)$, $G_n$ provides a test for $\mathrm{NSMC}_{r(n)}^{k(n)}(\mathbb{F})$, for all large enough $n$.*

**Theorem 2.7** *If Strengthened HH2 holds over $\mathbb{F}$, then there exists an efficiently computable multilinear $(O(n^{1/\sqrt{\log n}}), n)$-generator $\{G_n\}_{n \geq 1}$, such that for any $k(n) \in n^{O(1)}$ and $r(n) \in 2^{O(\sqrt{\log n})}$, $G_n$ provides a test for $\mathrm{NSMC}_{r(n)}^{k(n)}(\mathbb{F})$, for all large enough $n$.*

From this we will derive the following:

**Theorem 2.8** *If HH2 holds over $\mathbb{Q}$, then non-singular matrix completion over $\mathbb{Q}$ for matrices $M(x)$ of poly(n) size and with coefficients of poly(n) bits, where the individual degrees of $\det(M(x))$ are constant bounded, can be decided deterministically in time $2^{n^\epsilon}$, for any $\epsilon > 0$, provided $n$ is large enough.*

**Theorem 2.9** *If Strengthened HH2 holds over $\mathbb{Q}$, then non-singular matrix completion over $\mathbb{Q}$ for matrices $M(x)$ of poly(n) size and with coefficients of poly(n) bits, can be decided deterministically in time $2^{O(n^{1/\sqrt{\log n}} \log n)}$, under the promise that individual degrees of $\det(M(x))$ are bounded by $2^{O(\sqrt{\log n})}$.*

A central technical part of this paper is the following "Root Extraction Lemma" for skew circuits, which is of independent interest:

**Lemma 2.10** *Let $n, s$, and $m$ be integers with $s \geq n$. Let $P(x, y) \in \mathbb{F}[X_n, y]$ be a non-zero polynomial such that $L_{skew}(P) = s$. Let $f \in \mathbb{F}[X_n]$ be a polynomial with $\deg(f) = m$ such that $P(x, f(x)) \equiv 0$. Then $L_{skew}(f) \leq s \cdot 2^{O(\log^2 m)} r^{4 + \log m}$, where $r = \deg_y(P)$.*

Finally, we also prove the following randomness-to-hardness results:

**Theorem 2.11** *If for some $0 < \epsilon < 1$, there exists an efficiently computable multilinear $(\lceil n^\epsilon \rceil, n)$-generator $\{G_n\}_{n \geq 1}$, such that for any $k(n) \in n^{O(1)}$, $G_n$ provides a test for* $\mathrm{NSMC}_1^{k(n)}(\mathbb{F})$, *for all large enough n, then HH2 holds over* $\mathbb{F}$.

**Theorem 2.12** *If there exists an efficiently computable multilinear* $(O(n^{1/\sqrt{\log n}}), n)$-*generator* $\{G_n\}_{n\geq 1}$, *such that for any* $k(n) \in n^{O(1)}$, $G_n$ *provides a test for* $\mathrm{NSMC}_1^{k(n)}(\mathbb{F})$, *for all large enough n, then Strengthened HH2 holds over* $\mathbb{F}$.

Theorem 2.12 & Theorem 2.7 and Theorem 2.11 & Theorem 2.6 provide us with characterizations, which we summarize as follows:

**Corollary 2.13**

1. *HH2 holds over* $\mathbb{F}$ *if and only if there exists an efficiently computable multilinear* $(\lceil n^\epsilon \rceil, n)$-*generator* $\{G_n\}_{n\geq 1}$, *for some* $0 < \epsilon < 1$, *such that for all* $k(n) \in n^{O(1)}$, $G_n$ *provides a test for* $\mathrm{NSMC}_1^{k(n)}(\mathbb{F})$, *for all large enough n.*

2. *Strengthened HH2 holds over* $\mathbb{F}$ *if and only if there exists an efficiently computable multilinear* $(O(n^{1/\sqrt{\log n}}), n)$-*generator* $\{G_n\}_{n\geq 1}$, *such that for all* $k(n) \in n^{O(1)}$, $G_n$ *provides a test for* $\mathrm{NSMC}_1^{k(n)}(\mathbb{F})$, *for all large enough n.*

## 2.1 Comparison to Other Work

Part (i) of Theorem 7.7 in [KI04] can be phrased as follows:

**Theorem 2.14** *Assume HH1 holds over* $\mathbb{Q}$. *Let* $C$ *be a poly*$(n)$-*size arithmetical circuit over* $\mathbb{Z}$ *computing an n-variate polynomial* $f_n$ *of total degree poly*$(n)$ *and maximum coefficient size at most poly*$(n)$. *Testing whether* $f_n \equiv 0$ *can be done deterministically in time* $2^{n^\epsilon}$, *for any* $\epsilon > 0$, *provided n is large enough.*

Theorem 2.8 matches the time bound of Theorem 2.14. Thus we have shown an important special case of CPIT, for which deterministic subexponential time algorithms exist under weaker assumptions than was known previously. Similarly, Theorem 2.9 matches the time bound implicitly provided by Theorem 7.7 in [KI04] under Strengthened HH1, but using a weaker assumption.

## 3 Preliminaries

For a polynomial $f$, $H_k(f)$ denotes the homogeneous part of degree $k$, and $H_{\leq k}(f) \triangleq \sum_{i=0}^{k} H_i(f)$.

An *algebraic branching program* (ABP) $\Phi$ over $\mathbb{F} \cup X_n$ is given by a directed acyclic graph $G$ with source node $s$ and sink node $t$. Edges of $G$ are labeled with elements of $X_n \cup \mathbb{F}$. The *weight* of a directed path in $\Phi$ is defined to be the product of the edge labels. The polynomial computed by $\Phi$ is defined to be the sum of weights over all directed $s, t$-paths. For the size of $\Phi$ we count the number of edges in $G$. For a polynomial $f$, $B(f)$ is the size of any smallest ABP computing $f$. This generalizes in the obvious way to multi-output ABPs, by have several sink nodes $t_1, t_2, \ldots, t_m$. One easily proves the following proposition:

**Proposition 3.1** $L_{skew}(f) = \Theta(B(f))$.

We will use this to switch freely between skew circuits and ABPs. The latter model gives us some convenience. For example, for ABPs it is easy to see that if $f(x_1, x_2, \ldots, x_n)$ is computed an ABP $A$ of size $s_A$, and $g$ is computed by an ABP $B$ of size $s_B$, then $f(g, x_2, \ldots, x_n)$ can be computed by an ABP of size $O(s_A s_B)$. Indeed, simply replace each edge labeled with $x_1$ in $A$ with the $s, t$-dag given by $B$. Addition and multiplication of ABPs is done by parallel and series composition, respectively.

**Proposition 3.2** *Suppose $\Phi$ is a skew circuit of size $s$ computing $f \in \mathbb{F}[X_n]$. Then for any $i$, there exists a skew circuit of size $O(s \cdot i)$ computing $H_j(f)$ for all $0 \leq j \leq i$.*

**Proof.** This is achieved using the standard homogenization trick of keeping for each gate in $\Phi$, $i$ many copies that compute the homogeneous components up to degree $i$. □

**Lemma 3.3 (cf. Lemma 2.4 in [DSY08])** *Suppose $P(x, y) \in \mathbb{F}[X_n, y]$ can be computed by a skew circuit over $\mathbb{F}$ of size $s$. Then for any $i$, $\frac{\partial^i P}{\partial^i y}$ can be computed by a skew circuit of size $O(r \cdot s)$, where $r = \deg_y(P)$.*

**Proof.** Let $C(x, y)$ be a skew circuit for $P$ of size $s$. We can compute $C_0(x), C_1(x), \ldots, C_r(x)$ with an $r+1$-output skew circuit of size $O(r \cdot s)$ by evaluating $C(x, a_i)$ at $r + 1$ distinct elements $a_1, a_2, \ldots, a_{r+1} \in \mathbb{F}$, and then use linear interpolation. Next we can compute $\frac{\partial^i P}{\partial^i y}$ by adding $O(r^2)$ many gates. Since $r \leq s$, the lemma follows. □

**Lemma 3.4 (Lemma 2.1 in [Alo99])** *Let $f \in \mathbb{F}[X_n]$ be a non-zero polynomial such that the degree of $f$ in $x_i$ is bounded by $r_i$, and let $S_i \subseteq \mathbb{F}$ be of size at least $r_i + 1$, for all $i \in [n]$. Then there exists $(s_1, s_2, \ldots, s_n) \in S_1 \times S_2 \times \ldots \times S_n$ with $f(s_1, s_2, \ldots, s_n) \neq 0$.*

**Lemma 3.5 (Nisan-Wigderson Design [NW94])** *Let $n, m$ be integers with $n < 2^m$. There exists a family of sets $S_1, S_2, \ldots, S_n \subseteq [\ell]$, such that (1) $\ell = O(m^2 / \log n)$, (2) For each $i$, $|S_i| = m$, and (3) For every $i \neq j$, $|S_i \cap S_j| \leq \log n$. Furthermore, the above family of sets can be computed deterministically in time $\text{poly}(n, 2^\ell)$.*

Berkowitz [Ber84] observes that Samuelson's algorithm [Sam42] for computing the characteristic polynomial, does not use divisions and can be implement in NC$^2$ (Also see [MV97]). From this one derives the following statement, sufficient for our purpose:

**Proposition 3.6** *The determinant of an $n \times n$ matrix $M$ with integer entries of at most $m$ bits each can be computed in time $\text{poly}(n, m)$.*

# 4 Root Extraction within the Skew Circuit Model

We start with the observation that Theorem 3.1 in [DSY08] can be modified into the following lemma. A proof sketch is provided in Appendix A.

**Lemma 4.1** *Let $n, s$, and $m$ be integers with $s \geq n$. Let $P(x, y) \in \mathbb{F}[X_n, y]$ be a non-zero polynomial with $s = L_{skew}(P)$. Let $f \in \mathbb{F}[X_n]$ be a polynomial with $\deg(f) = m$ such that $P(x, f(x)) \equiv 0$. Then $L_{skew}(f) = O(s \cdot rm^{r+1})$, where $r = \deg_y(P)$.*

Comparing this with the $s \cdot 2^{O(\log^2 m)} r^{4+\log m}$ bound of Lemma 2.10, which can be bound by $s \cdot m^{O(\log m + \log r)}$, we see that we get a significant improvement for any $m << 2^r$.

Let us briefly indicate the idea behind the proof of Lemma 2.10. Similar as was done in [DSY08], we want to to approximate $f$ up to some degree $k$, i.e. find a polynomial $g$ with $H_{\leq k}(f) = H_{\leq k}(g)$. In [DSY08] this is done in increments of $k$ by one. This will not be good enough for our purpose. Due to the nature of the skew circuit model, typically any increment of $k$ requires duplication of previously constructed circuitry, leading to an overall exponential blowup by a factor of $2^m$. The solution is to aim for a faster *convergence rate* that doubles $k$ in stages. This way, one can keep circuit blow-up due to duplications more or less in check.

We now proceed with the proof of Lemma 2.10. In the following, for any polynomial $q$ the homogeneous component $H_t[q]$ will also be denoted by $q_t$.

**Lemma 4.2** *Let $P \in \mathbb{F}[X_n, y]$ be such that $\deg_y(P) = r$. Write $P = \sum_{i=0}^{r} C_i(x) y^i$, and let $P'(x, y) = \sum_{i=0}^{r} i C_i(x) y^{i-1}$. Let $f \in \mathbb{F}[X_n]$ be such that $P(x, f(x)) = 0$ and $P'(0, f(0)) = \xi_0 \neq 0$. Let $k \geq 1$ be an integer. Suppose $g \in \mathbb{F}[X_n]$ satisfies $H_{\leq k}[g] = H_{\leq k}[f]$. Then for any $1 \leq j \leq k$,*

$$f_{k+j} = g_{k+j} - \frac{1}{\xi_0}\left(P(x,g)_{k+j} + \sum_{i=1}^{j-1}(f_{k+i} - g_{k+i})P'(x,g)_{j-i}\right).$$

**Proof.** Let $h = (f_{k+1} - g_{k+1}) + \ldots + (f_{2k} - g_{2k})$. Then

$$\begin{aligned} 0 &= H_{\leq 2k}[P(x, f(x))] \\ &= H_{\leq 2k}[P(x, g + h)] \\ &= H_{\leq 2k}[\sum_{i=0}^{r} C_i(x)(g+h)^i] \\ &= H_{\leq 2k}[\sum_{i=0}^{r} C_i(x)(g^i + i \cdot g^{i-1} \cdot h)] \\ &= H_{\leq 2k}[P(x, g) + P'(x, g) \cdot h] \end{aligned}$$

Let $1 \leq j \leq k$ be given.

$$0 = P(x,g)_{k+j} + \sum_{i=1}^{j}(f_{k+i} - g_{k+i})P'(x,g)_{j-i}$$

$$= P(x,g)_{k+j} + (f_{k+j} - g_{k+j})P'(x,g)_0 + \sum_{i=1}^{j-1}(f_{k+i} - g_{k+i})P'(x,g)_{j-i}$$

Since $P'(x,g)_0 = P'(0,g(0)) = P'(0,f(0))$, the lemma follows. □

Applying the above lemma for $g = H_{\leq k}(f)$ yields the following corollary:

**Corollary 4.3** *Let $P \in \mathbb{F}[X_n, y]$ be such that $\deg_y(P) = r$. Write $P = \sum_{i=0}^{r} C_i(x)y^i$, and let $P'(x,y) = \sum_{i=0}^{r} iC_i(x)y^{i-1}$. Let $f \in \mathbb{F}[X_n]$ be such that $P(x, f(x)) = 0$ and $P'(0, f(0)) = \xi_0 \neq 0$. Let $k \geq 1$ be an integer. Then for any $1 \leq j \leq k$,*

$$f_{k+j} = -\frac{1}{\xi_0}\left(P(x,g)_{k+j} + \sum_{i=1}^{j-1} f_{k+i} \cdot P'(x,g)_{j-i}\right), \tag{1}$$

*where $g = H_{\leq k}[f]$.*

**Lemma 4.4** *Let $P \in \mathbb{F}[X_n, y]$ be such that $\deg_y(P) = r$. Write $P = \sum_{i=0}^{r} C_i(x)y^i$, and let $P'(x,y) = \sum_{i=0}^{r} iC_i(x)y^{i-1}$. Let $f \in \mathbb{F}[X_n]$ be such that $P(x, f(x)) = 0$ and $P'(0, f(0)) = \xi_0 \neq 0$. Let $k \geq 1$ be an integer. Let*

$$\mathcal{P} = \{P(x,g)_j : 1 \leq j \leq 2k\} \cup \{P'(x,g)_j : 1 \leq j \leq k-1\},$$

*where $g = H_{\leq k}[f]$. Suppose any polynomial in $\mathcal{P}$ can be computed by a single output ABP of size at most $B$. Then for any $1 \leq j \leq k$, there exist $(j+1)$-output ABP $\Phi_j$ computing $1, f_{k+1}, f_{k+2}, \ldots, f_{k+j}$ of size at most $2Bj^2$.*

**Proof.** We prove the lemma by induction on $j$. For $j = 1$, we see by Corollary 4.3 that $f_{k+1} = -\frac{1}{\xi_0}P(x,g)_{k+j}$. Hence we have an single output ABP computing $f_{k+1}$ of size at most $B$. This means we certainly can compute 1 and $f_{k+1}$ by means of a 2-output ABP of size at most $2B$.

Now suppose $1 < j < k$. By induction hypothesis we have $j$ output ABP $\Phi_{j-1}$ of size at most $2B(j-1)^2$ computing $1, f_{k+1}, f_{k+2}, \ldots, f_{k+j-1}$. The ABP $\Phi_j$ is constructed from $\Phi_{j-1}$ by first of all passing along all of $1, f_{k+1}, f_{k+2}, \ldots, f_{k+j-1}$ to the outputs. Then by drawing wires from each of these we can compute $f_{k+j}$ according to Equation (1). For this we use a new copy of a single output ABP computing some polynomial in $\mathcal{P}$ exactly $j$ times. A picture of the construction can be found in Appendix C. This construction can be implemented such that $size(\Phi_j) \leq size(\Phi_{j-1}) + jB + j + 1 \leq 2Bj^2$ (For this exact count we use that the cross wires are not actually needed, since we can identify nodes). □

**Lemma 4.5** *Let $n, s, r, m$ and be integers with $s \geq n$. Let $P \in \mathbb{F}[X_n, y]$ be a non-zero polynomial with $\deg_y(P) = r$. Write $P = \sum_{i=0}^{r} C_i(x)y^i$, and let $P'(x,y) = \sum_{i=0}^{r} iC_i(x)y^{i-1}$. Assume that both $P$ and $P'$ can be computed by skew circuits of size at most $s$ over $\mathbb{F}$. Let $f \in \mathbb{F}[X_n]$ be a polynomial with $\deg(f) = m$ such that $P(x, f(x)) \equiv 0$ and $P'(0, f(0)) \neq 0$. Then $f$ can be computed by a skew circuit of size at most $s \cdot 2^{O(\log^2 m)} r^{3+\log m}$.*

**Proof.** We compute $f$ in at most $\lceil \log m \rceil$ stages. At stage $i$ we construct an ABP $\Psi_i$ computing $H_{\leq 2^i}[f]$ of size $s_i$.

To start, for stage $i = 0$, since $H_{\leq 2^i}[f]$ is an affine linear form in $n$ variables, $\Psi_0$ can be constructed with $s_0 = O(n)$.

We now describe stage $i$, for $i > 0$. Let $g = H_{\leq 2^{i-1}}[f]$. In the previous stage an ABP $\Psi_{i-1}$ was constructed for $g$ of size $s_{i-1}$.

We claim $P(x,g)$ and $P'(x,g)$ can be computed by an ABP of size $O(rs_{i-1} + r^2 s)$. Namely, like in proof of Lemma 3.3, we have for any $i$, an ABP of size $O(rs)$ computing $C_i(x)$. Using $r$ copies of the ABP computing $g$ we can then compute $\sum_{i=0}^{r} C_i(x) g^i$ with size $O(rs_{i-1} + r^2 s)$. Similarly, for $P'(x,g)$.

Hence, by Proposition 3.2 and Proposition 3.1, for any $j \leq 2^i$, $P(x,g)_j$ can be computed by an ABP of size $O(2^i(rs_{i-1} + r^2 s))$. Similarly, for any $j \leq 2^{i-1}$, $P'(x,g)_j$ can be computed by an ABP of size $O(2^{i-1}(rs_{i-1} + r^2 s))$.

Therefore, we can apply Lemma 4.4 with $k = 2^{i-1}$ and $B := O(2^i(rs_{i-1} + r^2 s))$. This gives us an ABP $\Phi_{2k}$ computing $f_{k+1}, f_{k+2}, \ldots, f_{2k}$ of size at most $2Bk^2$. Combining $\Psi_k$ and $\Phi_{2k}$ to add all components of $f$ gives us the ABP $\Psi_{2k}$ computing $H_{\leq 2k}[f]$ of size $O(2^{3i}(rs_{i-1} + r^2 s) + s_{i-1})$. We can thus bound $s_i \leq \alpha r 2^{3i} \cdot (s_{i-1} + rs)$, for some absolute constant $\alpha > 1$. From this, one gets that $s_i \leq s \cdot \beta^{i^2+1} r^{i+2}$, for some absolute constant $\beta > 1$.

Taking $i = \lceil \log m \rceil$, we see there exists an ABP computing $f$ with size bounded by $s \cdot 2^{O(\log^2 m)} r^{3+\log m}$. Applying Proposition 3.1 completes the proof. □

## 4.1 Proof of Lemma 2.10

Write $P = \sum_{i=0}^{r} C_i(x) y^i$ with $C_r(x) \not\equiv 0$. Let $P^i(x,y) = \frac{\partial^i P}{\partial^i y}$. Then $P^r(x,y) = r! \cdot C_r(x)$. Since the characteristic of $\mathbb{F}$ is zero, $r! \neq 0$, and hence $P^r(x, f(x)) \not\equiv 0$. By assumption, $P^0(x, f(x)) \equiv 0$. Let $i$ be the smallest number such that $P^i(x, f(x)) \not\equiv 0$. Then $0 < i \leq r$, and $P^{i-1}(x, f(x)) \equiv 0$. By Lemma 3.4 there exists $x_0 \in \mathbb{F}$ such that $P^i(x_0, f(x_0)) \neq 0$.

Let $Q(x,y) = P^{i-1}(x + x_0, y)$, and let $g = f(x + x_0)$. $Q$ is computable by a skew circuit of size $O(r \cdot s)$ by Lemma 3.3. Let $Q' = \frac{\partial Q}{\partial y}$. Observe $Q'(x,y) = P^i(x + x_0, y)$. $Q$ is a nonzero polynomial such that $Q(x, g(x)) = P^{i-1}(x + x_0, f(x + x_0)) \equiv 0$, and $Q'(0, g(0)) = P^i(x_0, f(x_0)) \neq 0$. We apply Lemma 4.5 and obtain a skew circuit $\Psi$ computing $g(x)$ of size $s \cdot 2^{O(\log^2 m)} r^{4+\log m}$. From this a skew circuit computing $f$ is obtained that is at most a constant factor larger by performing the substitution $x := x - x_0$ within $\Psi$. □

## 5 Constructing a Generator from a Hard Polynomial

With the "Root Extraction" Lemmas 2.10 and 4.1 proved, the following lemma follows by the technique of Lemma 7.6 in [KI04], which was also employed to prove Lemma 4.1 in [DSY08]. The proof can be found in Appendix B.

**Lemma 5.1** *Let $n, r$ and $s$ be integers, and let $g \in \mathbb{F}[X_n]$ be a non-zero polynomial with individual degrees bounded by $r$ with $L_{skew}(g) = s \geq n$. Let $m > \log n$ be an integer and let $S_1, S_2, \ldots, S_n \subseteq [\ell]$ be given by Lemma 3.5, so that $\ell = O(m^2 / \log n)$, $|S_i| = m$,* and $|S_i \cap S_j| \leq \log n$. Let $f \in \mathbb{F}[z_1, z_2, \ldots, z_m]$ be a multilinear polynomial such that $g(f(y|_{S_1}), f(y|_{S_2}), \ldots, f(y|_{S_n})) \equiv 0$. Then $L_{skew}(f) \leq sn \cdot \min(2^{c_1(\log^2 m)} r^{4+\log m}, c_2 \cdot rm^{r+1})$, for absolute constants $c_1, c_2 > 1$.

## 5.1 Proof of Theorem 2.6 and 2.7

**Proof.** We first consider Theorem 2.7. Suppose $\{f_m\}$ is an explicit multilinear family with $dc(f_m) = m^{\omega(\log m)}$. Consider some large enough $n$. Set $m = \lceil 2^{\frac{1}{2}\sqrt{\log n}} \rceil$. Construct the Nisan-Wigderson design $S_1, S_2, \ldots, S_n$ as in Lemma 5.1 with $\ell(n) = O(m^2/\log n) = O(n^{1/\sqrt{\log n}})$. We claim the required $(\ell(n), n)$-generator $G_n$ can given by

$$G_n(y_1, y_2, \ldots, y_{\ell(n)}) \triangleq (f_m(y|_{S_1}), f_m(y|_{S_2}), \ldots, f_m(y|_{S_n})),$$

To verify this, consider any $k(n) \in n^{O(1)}$ and $r(n) \in 2^{O(\sqrt{\log n})}$, and arbitrary $k(n) \times k(n)$ matrix $M(x)$ with entries in $\mathcal{A}_\mathbb{F}(X_n)$. Let $g = \det(M(x))$. Assume the individual degrees of $g$ are bounded by $r(n) = poly(m)$. Observe it suffices to verify that if $g \not\equiv 0$, then $\det(M(G_n(y))) \not\equiv 0$. Due to [MV97], we know $g$ has a skew circuit over $\mathbb{F}$ of size at most $O(n \cdot k(n)^6) \leq n^d$, for some constant $d$ (provided $n$ is large enough). Hence by Lemma 5.1, if $\det(M(G_n(y))) \equiv 0$, we obtain a skew circuit over $\mathbb{F}$ for $f_m$ of size at most $n^{d+1} \cdot 2^{c_1(\log^2 m)} r(n)^{4+\log m} \leq 2^{4(d+1)\log^2 m} \cdot 2^{c_1(\log^2 m)} r(n)^{4+\log m}$. Since $r(n) = poly(m)$ and $n$ is assumed to be large enough, this contradicts the hardness of $f_m$. (Here we use $dc(f_m) = O(L_{skew}(f_m))$).

For Theorem 2.6 one argues similarly, but with $m := \lceil n^\epsilon \rceil$. We bound the size of the skew circuit for $f_m$ by $c_2 n^{d+1} \cdot r(n) m^{r(n)+1} \leq c_2 r(n) m^{(d+1)/\epsilon + r(n)+1}$. This contradicts the hardness of $f_m$, assuming $dc(f_m) = m^{\omega(1)}$, for any constant $0 < \epsilon < 1$ and $r(n) = O(1)$, provided $n$ is large enough.

We now check that in any of the above cases, $\{G_n\}_{n \geq 1}$ is efficiently computable. Given $(i, n, e)$, where $e \in \{0, 1\}^{\ell(n)}$, one first constructs the sets $S_1, S_2, \ldots, S_n$. This can be done deterministically in time $2^{O(\ell(n))}$ by Lemma 3.5. Then if for some $j \notin S_i$, $e_j = 1$, return zero. Otherwise, let $c = e|_{S_i}$. Return the coefficient of the monomial $x_1^{c_1} x_2^{c_2} \ldots x_m^{c_m}$ of $f_m$. Since $f_m$ is explicit, this coefficient can be computed deterministically in time $2^{O(m)}$. Hence the total deterministic time is bounded by $2^{O(\ell(n))}$. $\square$

**Remark 5.2** *From the above we see an $(\lceil n^\epsilon \rceil, n)$-generator for $\text{NSMC}_{r(n)}^{poly(n)}(\mathbb{F})$ can be obtained by assuming $dc(f_m) = m^{\omega(r(m^{1/\epsilon}))}$. For example, assuming $dc(f_m) = m^{\omega(\log \log m)}$ yields an $(\lceil n^\epsilon \rceil, n)$-generator for $\text{NSMC}_{\log \log n}^{poly(n)}(\mathbb{F})$, for any $0 < \epsilon < 1$. This observation is useful for $r(n) = o(\log n)$. Once $r(n) = \Theta(\log n)$, we known we are working under Strengthened HH2, which implies HH1, and we obtain an $(\lceil n^\epsilon \rceil, n)$-generator for CPIT from Theorem 2.14.*

## 6 Using the Generator to decide NSMC($\mathbb{Q}$) Deterministically

**Theorem 6.1** *Let $\ell(n)$ and $r(n)$ be functions of type $\mathbb{N} \to \mathbb{N}$ such that $\log n < \ell(n) < n$, for all large enough $n$. If there exists an efficiently computable multilinear $(\ell(n), n)$-generator $\{G_n\}_{n \geq 1}$, such that for any $p(n) \in n^{O(1)}$, $G_n$ provides a test for for $\text{NSMC}_{r(n)}^{p(n)}(\mathbb{Q})$, for all large* enough $n$, then for any $k(n) \in n^{O(1)}$, $\mathrm{NSMC}^{k(n)}_{r(n)}(\mathbb{Q})$ can be decided deterministically in time $2^{O(\ell(n)\log n + \ell(n)\log r(n))}$, provided coefficients of the input matrix have bit size $n^{O(1)}$.

**Proof.** Say $G_n$ is defined over variables $z_1, z_2, \ldots, z_{\ell(n)}$. Consider an arbitrary matrix $M$ of size $k(n)$, with entries in $\mathcal{A}_{\mathbb{Q}}(X_n)$, where coefficients have bit size $n^{O(1)}$, and with individual degrees of $\det(M(x))$ bounded by $r(n)$. We assume wlog. that entries of $M$ are in $\mathcal{A}_{\mathbb{Z}}(X_n)$, since we can multiply out all denominators and still leave bit sizes bounded by $n^{O(1)}$.

For large enough $n$, by Definition 2.5, $(\exists a \in \mathbb{Q}^n), \det M(a) \neq 0$ iff $(\exists b \in \mathbb{Q}^{\ell(n)}), \det M(G_n(b)) \neq 0$. Let $m = \ell(n)$. We have that $(\exists b \in \mathbb{Q}^m), \det M(G_n(b)) \neq 0$ if and only if $h := \det M(G_n(z)) \not\equiv 0$. Individual degrees of $h$ are at most $nr(n)$. By Lemma 3.4, if $h \not\equiv 0$, then for some $b \in V^m$, $h(b) \neq 0$, where $V = \{0, 1, \ldots, nr(n)\}$. Hence we can use the following test, for any $n$ larger than some fixed threshold depending on $k$:

**Algorithm** Test (input : an instance $M(x)$ of $\mathrm{NSMC}^{k(n)}_{r(n)}(\mathbb{Z})$)

1. Let $V = \{0, 1, \ldots, nr(n)\}$.

2. For all $b \in V^{\ell(n)}$, compute $v_b := \det(M(G_n(b)))$.

3. If for all $b \in V^{\ell(n)}$, $v_b = 0$, then **Reject** else **Accept**.

If the above algorithm accepts, one also knowns a non-singular completion. Let us estimate the running time. Since $G_n$ is efficiently computable, for any $b \in V^{\ell(n)}$, $G_n(b)_j$ can be computed in time $2^{O(m)}$. Each entry of $N := M(G_n(b))$ is an integer computable in time $2^{O(m)}$. By Proposition 3.6, $\det(N)$ is computable in time $poly(k(n), 2^{O(m)}) = 2^{O(m)}$. Hence the total time is bounded by $2^{O(m)} \cdot (nr(n)+1)^m = 2^{O(m\log n + m\log r(n))}$. $\square$

Using Theorem 6.1, the proofs of Theorem 2.8 and Theorem 2.9 immediately follow from Theorem 2.6 and Theorem 2.7, respectively.

## 7 Constructing a Hard Polynomial from a Generator

Let $\delta > 0$. We say a function $\ell : \mathbb{R}_{>0} \to \mathbb{R}_{>0}$ is $\delta$-*nice* if 1) $\ell$ is monotone increasing, 2) $\ell(t)^{1+\delta} < t$ and $|\ell(t+1)^{1+\delta} - \ell(t)^{1+\delta}| \leq 1$, for all large enough $t$, and 3) for all large enough $N$, given $N$ in unary, we can[2] compute an $n$ such that $N = \lceil \ell(n)^{1+\delta} \rceil$ deterministically in time $2^{O(N)}$.

**Theorem 7.1** *Let $\delta > 0$, and let $\ell : \mathbb{R}_{>0} \to \mathbb{R}_{>0}$ be a $\delta$-nice function. Given any efficiently computable multilinear $(\lceil \ell(n) \rceil, n)$-generator $\{G_n\}_{n \geq 1}$, we can construct an explicit multilinear family $\{g_N\}_{N \geq 1}$, such that if for some integer $d > 0$, $G_n$ provides a test for $\mathrm{NSMC}^{n^d}_1(\mathbb{F})$ for all large enough $n$, then for all large enough $N$, $dc(g_N) > \ell^{-1}(N^{1/(1+\delta)})^d$, over the field $\mathbb{F}$.*

---
[2]Note: conditions 1) and 2) imply the $n$ in condition 3) always exists, provided $N$ is large enough.

**Proof.** Consider some large enough $N$. Let $n$ be such that $N = \lceil \ell(n)^{1+\delta} \rceil$ (such $n$ can be found in time $2^{O(N)}$). Let $m = \lceil \ell(n) \rceil$. We have that $N \leq n$. Let $V = \{1, 2, \ldots, N+1\} \subseteq \mathbb{F}$. Similarly[3] as in [Agr05], define the polynomial $g_N(x_1, x_2, \ldots, x_N) = \sum_{I \subseteq [1,N]} c_I \prod_{i \in I} x_i$, where $c_I$ is taken to be an *integer* nonzero solution of the following system of linear equations:

$$\sum_{I \subseteq [1,N]} c_I \prod_{i \in I} G_n(a_1, a_2, \ldots, a_m)_i = 0, \tag{2}$$

for all $a \in V^m$. These are $(N+1)^m$ equations in $2^N$ variables. Provided $n$ is large enough, $m \log(N+1) < N$, and hence there exists a nonzero solution over $\mathbb{F}$. The technical conditions placed on $\ell(t)$ ensure $g_N$ is defined for all large enough $N$. Below we will argue how to compute an integer solution within time $2^{O(N)}$, so that $g_N$ is explicit in the sense of Definition 2.2.

For purpose of contradiction, suppose that $\mathrm{dc}(g_N) \leq n^d$. Hence we can write $g_N = \det(M)$, where $M$ is an $n^d \times n^d$ matrix with entries in $\mathcal{A}_\mathbb{F}(X_N)$. The entries of $M$ are elements of $\mathcal{A}_\mathbb{F}(X_n)$, since $\mathcal{A}_\mathbb{F}(X_N) \subseteq \mathcal{A}_\mathbb{F}(X_n)$. Since $\mathbb{F}$ is an infinite field and $g_N \not\equiv 0$, there exists $a \in \mathbb{F}^n$ such that $\det(M(a)) = g_N(a_1, a_2, \ldots, a_N) \neq 0$. The individual degrees of $g_N$ are bounded by one. Hence, by Definition 2.5, there exists $b \in \mathbb{F}^m$ such that $g_N(G_n(b)_1, G_n(b)_2, \ldots, G_n(b)_N) = \det(M(G_n(b))) \neq 0$. This implies $h \not\equiv 0$, where $h(z) := g_N(G_n(z)_1, G_n(z)_2, \ldots, G_n(z)_N)$. Observe that individual degrees of $h$ are bounded by $N$. Hence by Lemma 3.4, there exists $b' \in V^m$ such that $h(b') \neq 0$, but this contradicts (2). Therefore $\mathrm{dc}(g_N) > n^d \geq \ell^{-1}(N^{1/(1+\delta)})^d$, for all large enough $N$.

We now argue how to obtain an integer solution to (2). Since $G_n$ is efficiently computable, we can compute any coefficient $G_n(a_1, a_2, \ldots, a_m)_i$ by summing over all $2^m$ monomials. This takes time $2^{O(m)}$. We write (2) as $Ax = 0$, for an $r \times 2^N$ matrix $A$, with integer coefficients of bit size $2^{O(m)}$ and $r = (N+1)^m$. To construct $A$ takes time $2^{O(N)}$.

First, we want to find an independent set $S$ of $rank(A)$ many rows of $A$, and then extend $S$ to an independent set of size $2^N$. Let $e_1, e_2, \ldots, e_{2^N}$ denote the standard basis row-vectors of $\mathbb{F}^{2^N}$. One can do this as follows:

1. let $v_i$ equal row $i$ of $A$, for $i \in [r]$, and let $v_{r+i} = e_i$, for $i \in [2^N]$.

2. let $S = \emptyset$

3. for $i = 1$ to $r + 2^N$

4.     let $S' = S \cup \{v_i\}$

5.     compute $\beta = \det(BB^T)$, where $B$ is the $|S'| \times 2^N$ matrix of rows in $S'$.

6.     if $\beta \neq 0$, then set $S = S'$

By the Binet-Cauchy Theorem, $\det(BB^T) = \sum_{I \subseteq 2^N, |I| = |S'|} [\det(B_I)]^2$, where $B_I$ is the $|S'| \times |S'|$ matrix consisting of the columns in $I$ of $B$. Hence $\beta \neq 0$ if and only if there exists a set $I$ of $|S'|$ independent columns in $B$. The latter holds if and only if $S'$ is an independent set.

---
[3]Agrawal [Agr05] works with a different notion of a generator, and does not demand integer coefficients for explicitness.

The above procedure therefore maintains the invariant that after execution of line 6, $S$ is an independent set with $\{v_1, \ldots, v_i\} \subseteq span(S)$ (We use the convention that $\emptyset$ is an independent set with $span(\emptyset) = \{0\}$). This implies that after the $r$th iteration, $S$ contains $rank(A)$ many rows of $A$, and after the final iteration, $S$ is a basis.

Entries of $BB^T$ have bit size $2^{O(N)}$. By Proposition 3.6, $\det(BB^T)$ can be computed in time $2^{O(N)}$. Hence the above procedure takes time $2^{O(N)}$ in total.

Let $B$ be the matrix consisting of the rows in $S$ computed by the above procedure. $B$ is computable in time $2^{O(N)}$. Consider the adjugate $adj(B)$. It satisfies $B \cdot adj(B) = \det(B)I$. Hence we can pick a nonzero column from $adj(B)$ that is a solution to the original system (2). The entry $adj(B)_{ij} = (-1)^{i+j} M_{ji}$, where $M_{ij}$ is the determinant of the matrix $B$ with rows $i$ and $j$ removed. The latter is an integer, and by Proposition 3.6 it is computable in time $2^{O(N)}$. $\square$

One proves Theorem 2.11 using Theorem 7.1 with $\ell(t) = t^\epsilon$, and selecting a small $\delta > 0$ such that $\epsilon(1+\delta) \in \mathbb{Q} \cap (0,1)$. Then $\ell$ is $\delta$-nice. This yields an explicit multilinear family $\{g_N\}_{N \geq 1}$, such that for any $d$, for all large enough $N$, $dc(g_N) > N^{d/(\epsilon(1+\delta))}$. Hence $dc(g_N) = N^{\omega(1)}$.

For Theorem 2.12, assume wlog. $\{G_n\}_{n \geq 1}$ is an efficiently computable multilinear ($\lceil \ell(n) := c \cdot n^{1/\sqrt{\log n}} \rceil, n$)-generator, for constant $c \in \mathbb{Z}_{>0}$. Then $\ell^{-1}(n) = 2^{\log^2(n/c)}$, and $\ell$ is $\delta$-nice, for $\delta = 1$. Theorem 7.1 yields an explicit multilinear family $\{g_N\}_{N \geq 1}$, such that for any $d$, for all large enough $N$, $dc(g_N) > 2^{d \cdot \log^2(\frac{N^{1/2}}{c})}$. Hence $dc(g_N) = N^{\omega(\log N)}$.

# A  Proof Sketch of Lemma 4.1

Write $P = \sum_{i=0}^{r} C_i(x) y^i$. As in the proof of Lemma 2.10, one can reduce to the case where $\frac{\partial P}{\partial y}(0, (f(0)) \neq 0$. One can now follow the proof of Theorem 3.1 in [DSY08]. The only thing that we need to observe, is that in "Part III- Finding a circuit for $f(x)$" we can compute the polynomial $\tilde{Q}(z)$ using a skew circuit of size $O(m^r)$. By Lemma 3.3 we can compute each $C_i(x)$ using a skew circuit of size $O(s \cdot r)$ (This still holds regardless of the cost for reducing to the $\frac{\partial P}{\partial y}(0, (f(0)) \neq 0$-case). Hence $\tilde{Q}(C_0(x), C_1(x), \ldots, C_r(x))$ can be computed by a skew circuit of size $O(s \cdot rm^r)$ (See the remarks after Proposition 3.1). Hence using Proposition 3.2 we obtain a skew circuit for $f$ of size $O(s \cdot rm^{r+1})$. □

# B  Proof of Lemma 5.1

Define so-called hybrid polynomials:

$$H_0(x,y) = g(x_1, x_2, \ldots, x_n)$$
$$H_1(x,y) = g(f(y|S_1), x_2, \ldots, x_n)$$
$$H_2(x,y) = g(f(y|S_1), f(y|S_2), \ldots, x_n)$$
$$\vdots$$
$$H_n(x,y) = g(f(y|S_1), f(y|S_2), \ldots, f(y|S_n))$$

By assumption $H_0 \not\equiv 0$ and $H_n \equiv 0$. Hence there exists a smallest number $i$ such that $H_i \not\equiv 0$ and $H_{i+1} \equiv 0$.

We fix all variables $x_j$ for $i+2 \leq j \leq n$ and all $y_k$ not in $S_{i+1}$ to field constants by means of a substitution $\mathcal{S}$, such that after substitution still $H_i \not\equiv 0$. This is possible by Lemma 3.4, since $\mathbb{F}$ has infinite cardinality. Let $f'$ and $g'$ be $f$ and $g$, respectively, after substitution $\mathcal{S}$.

We now have

$$g'(f'(y|S_1 \cap S_{i+1}), f'(y|S_2 \cap S_{i+1}), \ldots, f'(y|S_i \cap S_{i+1}), x_{i+1}) \not\equiv 0 \qquad (3)$$
$$g'(f'(y|S_1 \cap S_{i+1}), f'(y|S_2 \cap S_{i+1}), \ldots, f'(y|S_i \cap S_{i+1}, f(y|S_{i+1})) \equiv 0 \qquad (4)$$

To simplify the discourse, rename $x_{i+1}$ by $w$ and the $y$-variables in $S_{i+1}$ by $z_1, z_2, \ldots, z_m$. Equations (3) and (4) then yield we have a polynomial $p$ such that

$$p(z_1, z_2, \ldots, z_m, w) \not\equiv 0$$
$$p(z_1, z_2, \ldots, z_m, f(z_1, z_2, \ldots, z_m)) \equiv 0$$

Let us argue that $p$ can be computed by an ABP of size $O(ns)$. Each $f'(y|S_j \cap S_i + 1)$ is a multilinear polynomial in at most $\log n$ variables, and hence can be computed by an ABP of size $O(n)$. $g'$ has an ABP of size at most $O(s)$. Hence $p$ can be computed by an ABP of size $O(ns)$.

By Proposition 3.1 we have that $L_{skew}(p) = O(ns)$. Since individual degrees of $g$ are bounded by $r$, $\deg_w(p) \leq r$. We apply Lemma 2.10 and Lemma 4.1 to conclude that $f$ can be computed by a skew circuit of size at most $sn \cdot \min(2^{c_1(\log^2 m)} r^{4+\log m}, c_2 \cdot rm^{r+1})$, for absolute constants $c_1, c_2 > 1$. $\square$

# C

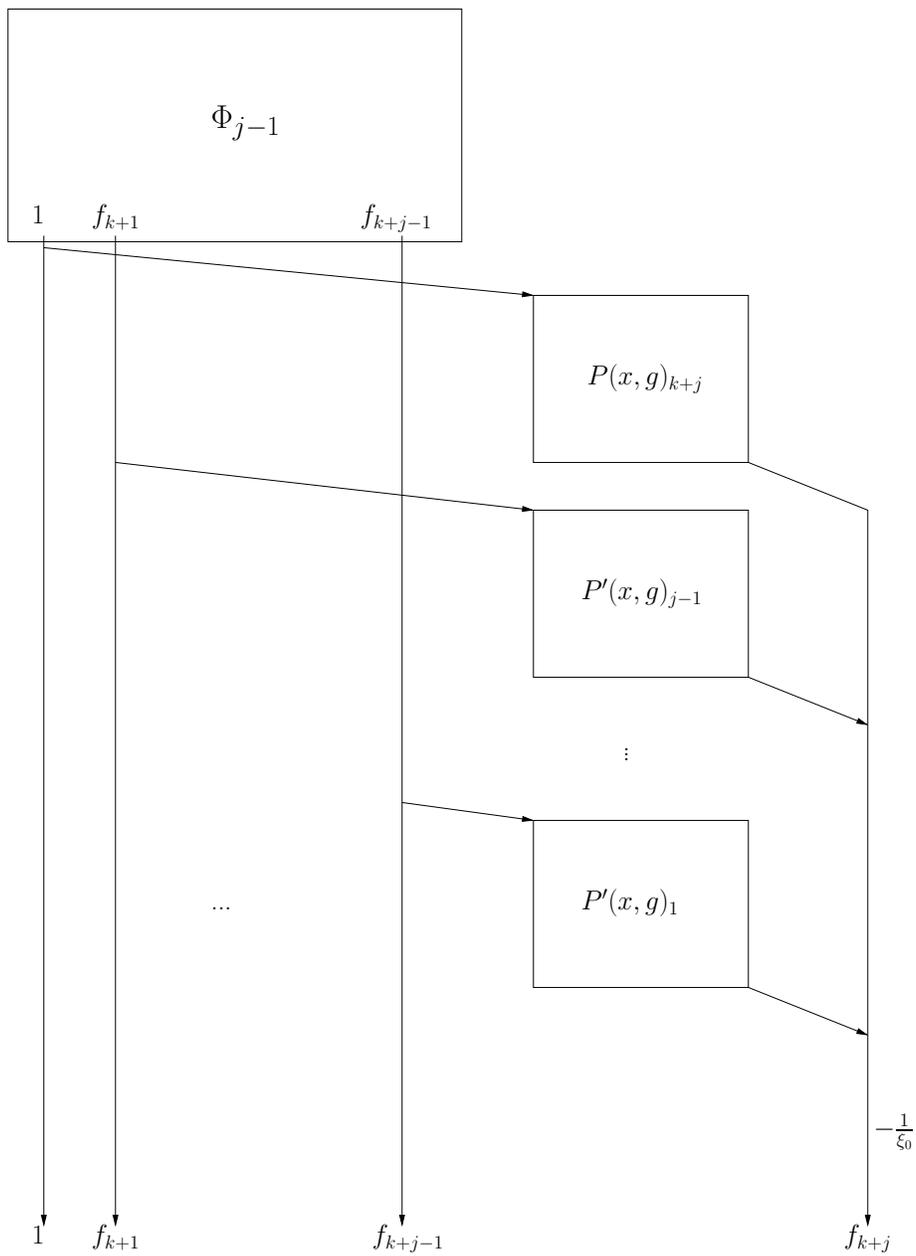

Figure 1: Construction of $\Phi_j$ in the proof of Lemma 4.4.